\documentclass[rnote]{aa}

\usepackage{graphicx}
\usepackage{txfonts}
\usepackage{url}

\newcommand{\etal}{et al.~}
\newcommand{\be}{\begin{equation}}
\newcommand{\ee}{\end{equation}}
\newcommand{\bea}{\begin{eqnarray}}
\newcommand{\eea}{\end{eqnarray}}
\newcommand{\HII}{\hbox{H\,{\sc ii}}}

\newcommand{\HI}{\hbox{H\,{\sc i}}~}

\newcommand{\Msun}{~M$_{\odot}$~}

\newcommand{\gtsim}{\stackrel >{_\sim}}
\newcommand{\ltsim}{\stackrel <{_\sim}}

\begin{document}

\title{A compact steep spectrum radio source in NGC~1977}

\author{D. Anish Roshi\inst{1,}\inst{2}
        \and
        Scott M. Ransom\inst{1}
       }

\institute{National Radio Astronomy Observatory 
              \thanks{The National Radio Astronomy Observatory is
               a facility of the National Science Foundation operated under cooperative
               agreement by Associated Universities, Inc.
              }, 
              Charlottesville, VA 22903-4608 \\
              \email{aroshi@nrao.edu, sransom@nrao.edu}
         \and
           National Radio Astronomy Observatory,
           Green Bank, WV 24944 \\ 
             }

   \date{}

\abstract
{ A compact steep spectrum radio source (J0535$-$0452) is located in the sky coincident 
with a bright optical rim in the \HII\ region NGC~1977. 
J0535$-$0452 is observed to be $\leq 100$~mas in angular size at 8.44 GHz. The spectrum 
for the radio source is steep and straight with a spectral index of $-1.3$ 
between 330 and 8440 MHz. No 2 $\mu m$ IR counter part for the source is detected.
These characteristics indicate that the source may be either a rare high redshift radio
galaxy or a millisecond pulsar (MSP).} 
{To investigate whether the steep spectrum source is a millisecond pulsar.
The optical rim is believed to be the interface between the \HII\ region
and the adjacent molecular cloud. If the compact source is a millisecond pulsar,
it would have eluded detection in previous pulsar surveys because of the extreme
scattering due to the \HII\ region--molecular cloud interface. 
}
{The limits obtained on the angular broadening along with the distance to the
scattering screen are used to estimate the pulse broadening. The pulse broadening
is shown to be less than a few msec at frequencies $\gtsim$ 5 GHz. We therefore
searched for pulsed emission from J0535$-$0452 at 14.8 and 4.8 GHz with the 
Green Bank Telescope (GBT).} 
{No pulsed emission is detected to 55 and 30 $\mu$Jy level at 4.8 and 14.8 GHz.
Based on the parameter space explored by our pulsar search algorithm, we
conclude that, if J0535$-$0452 is a pulsar, then it could only be a binary MSP 
of orbital period $\ltsim$ 5 hrs.}
{}

\keywords{Stars: pulsars -- ISM: \HII\ regions -- Scattering}

\maketitle

\section{Introduction}

NGC1977 is an \HII\ region located at the northern edge of the Orion molecular
cloud at a distance of about 0.5 kpc from the Sun. The region has
an emission measure of 5 $\times$ 10$^3$ pc cm$^{-6}$ and is believed to be
an example of an interface between the \HII\ region and the adjacent molecular
cloud (Shaver \& Goss 1970). A 6-cm VLA image of the \HII\ region made by 
Kutner \etal (1985) shows a strong peak that coincides with the brightest 
part of the optical bright rim. They investigated whether the 
radio peak is due to a compact \HII\ region or dense clump of gas ionized by
the central star in NGC1977. Both these possibilities were ruled out based 
on the expected radio and IR emission from such ionized gas.
Subrahmanyan, Goss, \& Malin (2001) imaged NGC~1977 with the VLA at 330, 
1420, 4860 and 8440 MHz. They detected the bright radio peak as an 
unresolved object designated as J0535$-$0452. The estimated flux density of the compact object 
is 320, 50, 11.4 and 5.2 mJy at 330, 1420, 4860 and 8440 MHz respectively.  
The spectrum of the object is straight, with no turn over down to 330 MHz. 
The spectral index is $-$1.3 over this range of frequencies.
The 8440 MHz observations are used to obtain an upper-limit
on the source size of $\sim$ 100 mas. The implied  brightness temperature is
$7 \times 10^8$~K. Subrahmanyan \etal\ (2001) have concluded that the object is a 
non-thermal source.

Kutner \etal\ (1985) argued that the non-thermal source could be an 
extragalactic background source located behind NGC1977. Compact steep spectrum
extragalactic sources with spectral index close to $-$1.3 are typically 
high redshift objects ($\gtsim$ 2; \nocite{md08}Miley \& De Breuck 2008). 
For these redshifts, the size of J0535$-$0452 is $\ltsim$ 780 pc, assuming standard
cosmological parameters. Compact steep spectrum sources of this size typically show
spectral turn over near a few GHz, which is not the case for J0535$-$0452.
Thus if J0535$-$0452 is an extragalactic source, then it should be one
of the rare steep-spectrum, high redshift radio galaxies (for example, 
see \nocite{ketal00}Kaplan \etal\ 2000).  

Another possibility is that the compact source
in NGC1977 is a pulsar. The spectral indices of both normal and 
millisecond pulsar are in the range $-$1 to $-$2.5 with a mean 
value of $-$1.65 (\nocite{ketal98}Karmer et al. 1998). 
Thus the compactness and spectral index of J0535$-$0452 may suggest that 
it is a pulsar. An earlier attempt to detect pulsed emission near 330 MHz was not successful
(Ramachandran, R. 1995; unpublished). However, 
J0535$-$0452 would have eluded detection in earlier pulsar 
searches due to extreme scattering. We show in Section~\ref{onobs}
that the best frequencies to search for pulsed emission from this pulsar 
are $\ge$ 5 GHz depending on the pulsar period. We used the GBT
to search for pulsed emission at frequencies near 4.8 and 14.8 GHz, 
but did not detect pulsed emission. The
observations and results are given in Section~\ref{gbtobs} and \ref{result},
respectively. 

\section{On observing pulsed emission from J0535$-$0452}
\label{onobs}
\begin{figure*}
\centering
\includegraphics[width=\textwidth]{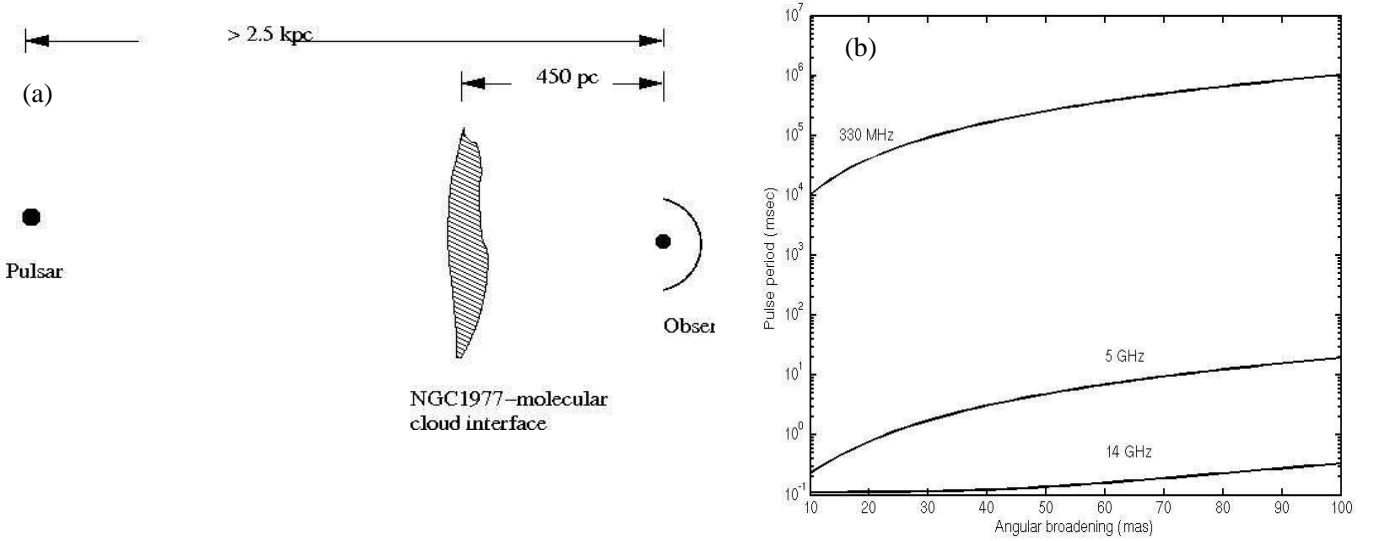}
\caption{(a)Schematic showing the scattering geometry. The \HII\ region--molecular cloud 
interface (scattering screen) is located at $\sim$ 450 pc from the observer and the 
distance to the compact source is $\ge$ 2.5 kpc.
(b) The expected pulse broadening for three frequencies (330 MHz, 5 and 14 GHz) 
is shown in the pulsar period--angular broadening parameter space. The curves are
the locus of points for pulse broadening equal to the pulsar period.
For each frequency the parameter space above the curve can be probed by the pulsar search 
observation. The pulse broadening estimation includes contribution from an assumed intrinsic
duty cycle of 10\% of the pulsar period and broadening due to DM.} For these calculation,
the smallest pulsar period is taken as 1 msec. The range of angular broadening 
is constrained by observations between 10 and 100 mas. For all frequencies $\ltsim$ 14 GHz,
the pulse broadening is dominated by scattering. At 14 GHz, for angular broadening $\ltsim$ 60 mas,
the pulse broadening is limited by the assumed intrinsic duty cycle.
\label{fig1}

\end{figure*}

An attempt was made in 2003 to image J0535$-$0452 using the VLBA at 8.4 GHz 
(Roshi \etal\ 2003 unpublished). The source was not detected in
this observation. A possible interpretation of the non-detection in the VLBA
observations is that the source would have angular broadened due
to scattering in the \HII\ region -- molecular cloud interface. Assuming
that the source is scatter broadened, a lower limit can be obtained for angular 
broadening which turns out to be $\sim$ 10 mas.  
If J0535$-$0452 is a pulsar, then the pulsed emission will be broadened due 
to scattering in the \HII\ region -- molecular cloud interface. 
We estimate the pulse broadening
using the measured limits on angular broadening of J0535$-$0452 along with 
the source-scattering screen geometry shown in Fig~\ref{fig1}a. The distance to the 
compact object, $D_{pul}$, as measured from \HI\ absorption studies is
$\ge$ 2.5 kpc (Subrahmanyan \etal\ 2001) and the distance
to NGC1977, $D_{hii}$, is $\sim$ 450 pc (Genzel \etal\ 1981). 
The $e^{-1}$ pulse broadening time in sec due to scattering is obtained using the equation 
(Cordes \& Lazio 1997) 
\be
t_s = \frac{D_{hii}}{D_{pul} - D_{hii}} \times
         \frac{D_{pul}\; \theta_{s,8.4GHz}^2}{8 \; \mbox{ln}(2)\; c} \times \left(\frac{f_{GHz}}{8.44}\right)^{-4},
\ee
where $c$ is the velocity of light in cm s$^{-1}$, the units of distances are in cm, 
$f_{GHz}$ is the observing frequency in GHz and $\theta_{s,8.4GHz}$ is the angular 
broadening at 8.4 GHz in rad. The pulse broadening time is considered 
to scale with the 4$^{th}$ power of 
frequency, appropriate for strong scattering. $\theta_{s,8.4GHz}$
range between 10 and 100 mas, which are obtained from the VLA (Subrahmanyan \etal\ 2001)
and VLBA observations near 8.4 GHz (Roshi \etal\ 2003 unpublished).
In Fig~\ref{fig1}b we plot the estimated pulse broadening time 
for three observing frequencies (330 MHz, 5 and 14 GHz) in the pulse period--
angular broadening parameter space.  The pulse broadening time in this plot
also includes an assumed intrinsic duty cycle of 10\% of the pulsar period
(minimum pulsar period is taken as 1 msec). The spectral resolution
used for the calculation is 0.78 MHz. For a given dispersion measure (DM),
this finite spectral resolution produces temporal broadening, which is also
added to the plotted pulse broadening. Any temporal broadening due to DM griding
in the pulsar search algorithm is neglected.

Continuum observations toward J0535$-$0452 can be used to estimate limits on the DM.
The measured spectrum of the compact source does not show any turnover at 330 MHz
due to the continuum optical depth of the \HII-region--molecular cloud interface.
This fact is used to obtain an upper limit on the emission measure (EM) of $\sim 10^4$ cm$^{-6}$ pc
for an assumed ionized gas temperature of 8000 K and $\tau_c \sim$ 0.05 at 330 MHz.
The 3 sigma uncertainty of the 330 MHz map of NGC1977 region (Subrahmanyan \etal\ 2001)
is used to infer an upper limit on $\tau_c$ of $\sim$ 0.05. A DM 
corresponding to a given EM can be obtained as $DM = \sqrt{EM\, \phi\, L}$,
where $\phi$ is the filling factor of the ionized gas in the intervening \HII\ region
and $L$ is the line of sight extent of the \HII\ region. Taking $\phi \sim 1$ and
$L \sim 10$ pc, a good fraction of the size of the Orion molecular cloud, the DM
we get is $\sim$ 300 pc cm$^{-3}$. Contribution to DM due to the distributed ionized gas
has to be added to the above estimate. NE2001 model (Cordes \& Lazio 2002) provides
a total DM of $\sim$ 50 pc cm$^{-3}$ in the direction of NGC1977. Thus the upper
limit obtained on DM is $\sim$ 350 pc cm$^{-3}$. For estimating pulse broadening we used
DM values up to 650 pc cm$^{-3}$.  
 
The result of the pulse broadening calculation shows that for all frequencies $\ltsim$ 14 GHz
the pulse broadening is dominated by scattering (see Fig~\ref{fig1}b). At 14 GHz, for $\theta_{s,8.4GHz} < 60$ mas,
pulse broadening plotted in Fig~\ref{fig1}b is limited by the assumed intrinsic
duty cycle. For each observing frequency indicated 
in Fig~\ref{fig1}b pulsed emission can be detected in the parameter space 
above the corresponding curve. 

The estimated continuum flux density at 14.0 GHz is 2.5 mJy.
If the observed continuum emission is the mean flux density of the pulsar, then the
pulsar can be detected at frequencies 5 and 14 GHz (depending on the pulsar period) 
with signal-to-noise ratio $>$ 100 in 1 hr of observing time with the GBT. 
Thus, as seen in Fig.~\ref{fig1}b,
such an observation will be sensitive to both MSP and ordinary pulsars.
Note that earlier searches would not have detected pulsars for the measured
angular broadening limits (see Fig~\ref{fig1}b).

\section{The GBT Observations and Data Analysis}
\label{gbtobs}

The GBT observations on J0535$-$0452 were made at 4.8 and 14.8 GHz
on 15 May 2011. The GUPPI backend with 800 MHz bandwidth, 1.6 MHz 
spectral resolution and
41 $\mu$sec time resolution was used for the observations. 
Flux density calibration was done using the source 3C161 for both
frequencies. The flux densities of 3C161 were taken to be 6.7 and
2 Jy at 4.8 and 14.8 GHz (Ott \etal\ 1994). The measured telescope
gains were 2 and 1.9 K/Jy and system temperatures were 18.5 and 26 K 
at 4.8 and 14.8 GHz respectively.
Reference pointing was done on B0540-0415 at 14.8 GHz
and on 3C161 at 4.8 GHz. The total on-source observing time at 14.8 GHz
is 27 minutes and that at 4.8 GHz is 29 minutes. 
The data processing was performed using PRESTO\footnote{
\url{http://www.nrao.edu/~sransom/presto}} (Ransom, Eikenberry \& Middleditch 2002). 
A DM range of 0 to 720 pc cm$^{-3}$ was searched with full time resolution. 
The value of 720 pc cm$^{-3}$ is about two times larger than the upper limit on 
DM estimated in Section~\ref{onobs}.  In addition we searched for pulsar 
acceleration in the range 0 to 2.1 $\times 10^4 \times$ $P$
m s$^{-2}$, where $P$ is the period of the pulsar in sec.

\section{Results and Conclusion}
\label{result}

We did not detect pulsed emission from J0535$-$0452. The upper limit 
obtained for pulsed emission at 4.8 GHz is 55 $\mu$Jy for a pulsar period
of 10 msec and pulse broadening of 7 msec corresponding to an angular 
broadening of 60 mas (see Fig~\ref{fig1}b). At 14.8 GHz, the upper limit
obtained is 30 $\mu$Jy for a pulsar period of 3 msec, typical for MSP,
assuming 10\% intrinsic duty cycle. 

Acceleration searches are most sensitive to binary pulsars with orbital period greater
than 10 times the observation time (\nocite{rce03}Ransom, Cordes \& Eikenberry 2003). 
This means that our data analysis will rule out compact sources in binary system 
with orbital period $\gtsim$ 5 hrs (assuming circular orbit). 
We have searched for accelerations up to 2.1 $\times 10^4 \times$ $P$ m s$^{-2}$.
For orbital period of $\sim$ 5 hrs, this upper limit rules out companion object
of mass $\ltsim$ 3 \Msun for an assumed pulsar period of 3 msec. If the companion 
objects is of higher mass (ie $>$ 3 \Msun) then for orbital periods $\gtsim$
5 hrs the object cannot be an ordinary star. This is because the orbital radius
becomes comparable to the size of the star. The companion object could be
a stellar mass black hole. However, we are not aware of any other observational
evidence for the presence of a stellar mass black hole in the direction of NGC1977. 
Thus the only possibility that remains is that J0535$-$0452 is a binary MSP with
short orbital period (ie $\ltsim$ 5 hrs).  

\begin{acknowledgements}
We thank the anonymous referee for the critical comments on the manuscript.
\end{acknowledgements}

\end{document}